\documentclass{article}
\usepackage{spconf}
\usepackage[utf8]{inputenc}
\usepackage[T1]{fontenc}

\usepackage{cite}

\usepackage{amsmath,amssymb,amsthm,bm,xfrac}
\usepackage{mathtools}

\usepackage{graphicx,subcaption}

\usepackage{xcolor}
\definecolor{comb_lapl}{rgb}{0,0.4470,0.7410}
\definecolor{rw_lapl}{rgb}{0.8500,0.3250,0.0980}
\definecolor{voro_lapl}{rgb}{0.9290,0.6940,0.1250}

\usepackage[inline]{enumitem}
\newlist{inlinelist}{enumerate*}{1}
\setlist*[inlinelist,1]{label=\roman*),itemjoin={{, }},itemjoin*={{, and }}}

\usepackage{hyperref}
\def\equationautorefname~#1\null{(#1)\null}

\usepackage{xspace}

\def\ie{\textit{i.e.}\xspace}
\def\eg{\textit{e.g.}\xspace}

\def\Q{\ensuremath{{\color{blue} \mathbf{Q}}}\xspace}
\def\Qinv{\ensuremath{{\color{blue} \mathbf{Q}^{-1}}}\xspace}
\def\Qsqrt{\ensuremath{{\color{blue} \mathbf{Q}^{\frac{1}{2}}}}\xspace}
\def\QSCsqrtinv{\ensuremath{{\color{blue} \mathbf{Q}_{\mathcal{S}^c}^{-\frac{1}{2}}}}\xspace}
\def\QS{\ensuremath{{\color{blue} \mathbf{Q}_{\mathcal{S}}}}\xspace}
\def\QSC{\ensuremath{{\color{blue} \mathbf{Q}_{\mathcal{S}^c}}}\xspace}
\def\QSinv{\ensuremath{{\color{blue} \mathbf{Q}_{\mathcal{S}}^{-1}}}\xspace}
\def\QSsqrt{\ensuremath{{\color{blue} \mathbf{Q}_{\mathcal{S}}^{\frac{1}{2}}}}\xspace}
\def\QSsqrtsm{\ensuremath{{\color{blue} \mathbf{Q}_{\mathcal{S}}^{1/2}}}\xspace}
\def\Qsqrtinv{\ensuremath{{\color{blue} \mathbf{Q}^{-\frac{1}{2}}}}\xspace}

\def\zerov{\ensuremath{\mathbf{0}}\xspace}
\def\onev{\ensuremath{\mathbf{1}}\xspace}

\def\ev{\ensuremath{\mathbf{e}}\xspace}

\def\nv{\ensuremath{\mathbf{n}}\xspace}
\def\tv{\ensuremath{\mathbf{t}}\xspace}
\def\uv{\ensuremath{\mathbf{u}}\xspace}
\def\xv{\ensuremath{\mathbf{x}}\xspace}
\def\yv{\ensuremath{\mathbf{y}}\xspace}
\def\zv{\ensuremath{\mathbf{z}}\xspace}

\def\phiv{\ensuremath{\bm{\phi}}\xspace}

\def\Cm{\ensuremath{\mathbf{C}}\xspace}
\def\Dm{\ensuremath{\mathbf{D}}\xspace}
\def\Em{\ensuremath{\mathbf{E}}\xspace}
\def\Fm{\ensuremath{\mathbf{F}}\xspace}
\def\Hm{\ensuremath{\mathbf{H}}\xspace}
\def\Im{\ensuremath{\mathbf{I}}\xspace}
\def\Km{\ensuremath{\mathbf{K}}\xspace}
\def\Lm{\ensuremath{\mathbf{L}}\xspace}
\def\Mm{\ensuremath{\mathbf{M}}\xspace}
\def\Um{\ensuremath{\mathbf{U}}\xspace}
\def\Wm{\ensuremath{\mathbf{W}}\xspace}
\def\Zm{\ensuremath{\mathbf{Z}}\xspace}

\DeclareMathOperator{\diag}{diag}
\DeclareMathOperator{\dist}{dist}
\DeclareMathOperator{\trace}{tr}
\DeclareMathOperator*{\argmin}{argmin}
\DeclareMathOperator*{\argmax}{argmax}

\newcommand*\xbar[1]{%
   \hbox{%
     \vbox{%
       \hrule height 0.5pt 
       \kern0.5ex
       \hbox{%
         \kern-0.1em
         \ensuremath{#1}%
         \kern-0.1em
       }%
     }%
   }%
}

\title{Graph Vertex Sampling with Arbitrary Graph Signal Hilbert Spaces}

\name{Benjamin Girault, Antonio Ortega, Shrikanth S. Narayayan
\thanks{The research is based upon work supported by the Office of the Director 
of National Intelligence (ODNI), Intelligence Advanced Research Projects 
Activity (IARPA), via IARPA Contract No 2017-17042800005. The views and conclusions contained herein are 
those of the authors and should not be interpreted as necessarily 
representing the official policies or endorsements, either expressed or 
implied, of the ODNI, IARPA, or the U.S. Government. The U.S. Government 
is authorized to reproduce and distribute reprints for Governmental 
purposes notwithstanding any copyright annotation thereon.}
\thanks{Work partially funded by NSF under grant CCF-1527874.}}
\address{Signal and Image Processing Institute, University of Southern California, Los Angeles, CA 90089}

\begin{document}
\ninept
\maketitle
\begin{abstract}
Graph vertex sampling set selection aims at selecting a set of vertices of a graph such that the space of graph signals that can be reconstructed exactly from those samples alone is maximal.
In this context, we propose to extend sampling set selection based on spectral proxies to arbitrary Hilbert spaces of graph signals.
Enabling arbitrary inner product of graph signals allows then to better account for vertex importance on the graph for a sampling adapted to the application.
We first state how the change of inner product impacts sampling set selection and reconstruction, and then apply it in the context of geometric graphs to highlight how choosing an alternative inner product matrix can help sampling set selection and reconstruction.
\end{abstract}

\section{Introduction}

Over the past few years, the field of \emph{graph signal processing} (GSP) has been building up a framework to analyze, process and transform graph supported data with the goal of understanding data supported by arbitrary discrete structures \cite{Shuman.SPMAG.2013,Ortega.PROCIEEE}.
In this paper, we are interested in the question of sampling set selection: Where on the graph do we perform measurements so as the dimension of the space of smooth signals we can reconstruct exactly is as high as possible?
While there have been a number of contributions studying this problem in a GSP context in the past few years \cite{Shomorony.GLOBALSIP.2014,Chen.TSP.2015,Anis.TSP.2016},  in all of them the space of graph signals is assumed to be equipped with the canonical inner product: the dot product.

Recently, we introduced a generalization of the graph Fourier transform to arbitrary Hilbert spaces of graph signals, allowing to equip that space with any inner product \cite{Girault.TSP.2018}.
We showed that this additional parameter (\ie, the inner product) allows better control over the importance of the nodes of the graph, with their relative importance influencing the design of a suitable graph Fourier transform.
For example, the random walk Laplacian approach based on the inner product whose matrix is the degree matrix \Dm, is shown to yield the most interesting results in spectral clustering \cite{vonLuxburg.SC.2007,Girault.TSP.2018}.
In image processing, bilateral filtering and non-local means are based on the inner product $\Im+\Dm$ and perform well in image denoising \cite{Milanfar.SPMAG.2013,Girault.TSP.2018}.
Still in image processing, a recent work shows that the inner product matrix can be used towards perceptually better compression by integrating the importance of each pixel towards the \emph{Structural SIMilarity} (SSIM) \cite{Lu.ICIPPRE.2020}.
Finally, in point cloud attribute compression, a region-adaptive approach can be seen as a hierarchical graph Fourier transform with specific graphs and inner products at each level \cite{deQueiroz.TIP.2016,Chou.TIP.2020,Pavez.ICIPPRE.2020}.

Our contribution is twofold.
First, in \autoref{sec:method}, we show how the sampling approach of \cite{Anis.TSP.2016} can be generalized to any inner product.
In particular, our generalization describes a method that can applied to many other GSP tools to extend them to  arbitrary Hilbert spaces, not just sampling set selection.
Our second contribution, in \autoref{sec:experiments}, is to experimentally study the impact of the inner product on sampling and reconstruction for geometric graphs and several inner product.
Doing so, we show that it has an impact on both theoretical worse case scenario error bounds and mean squared reconstruction error, and we give evidence that the Voronoi cell area inner product proposed in \cite{Girault.TSP.2018} is a good candidate.

\section{Sampling with Spectral Proxies}
\label{sec:method}

In this section, we describe how the graph vertex sampling set selection approach of \cite{Anis.TSP.2016} can be extended to any arbitrary Hilbert space of graph signals.
To that end, we use the graph Fourier transform adapted to arbitrary Hilbert spaces that we proposed in \cite{Girault.TSP.2018} to obtain the required notion of bandlimitedness.

\subsection{Background: Graph Fourier Transform in Arbitrary Hilbert Spaces}

Let $\mathcal{G}=(\mathcal{V},\mathcal{E},\Wm)$ be a graph with vertex set $|\mathcal{V}|=N$, edge set $\mathcal{E}$, and weight matrix \Wm such that $w_{ij}$ is the weight of the edge from vertex $i$ to vertex $j$.
$\mathcal{G}$ may be directed or undirected.
A graph signal \xv maps any vertex $i$ to a scalar value $\xv_i\in\mathbb{C}$.
We assume that there exists a non-negative graph signal variation operator $\Delta(\xv)\geq 0$ that depends on the edge set $\mathcal{E}$ and weight matrix \Wm.
In \cite{Girault.TSP.2018}, we proposed a generalization of the \emph{graph Fourier transform} (GFT) for any Hilbert space of graph signals using the variation operator $\Delta$.
We denote \Q the matrix of the graph signal Hilbert space inner product verifying $\langle\xv,\yv\rangle_\Q=\yv^*\Q\xv$ with $.^*$ the conjugate transpose operator.
Above, and in the rest of this paper, we color the inner product matrix \Q in blue to highlight where it impacts the sampling set selection approach of \cite{Anis.TSP.2016}.
In this paper, we further assume that the variation operator verifies $\Delta(\xv)=\xv^*\Mm\xv$ with $\Mm$ a non-negative Hermitian matrix.%
\footnote{Extending our current contribution to any $\Delta$ by considering the fundamental matrix of the graph $\Zm$ instead of $\Qinv\Mm$ is straightforward, but we choose not to do it here to better highlight how \Q changes the setting of \cite{Anis.TSP.2016}.}
We denote this graph Fourier transform the $(\Mm,\Q)$-GFT.

More precisely, the $(\Mm,\Q)$-GFT projects a graph signal onto an orthonormal%
\footnote{Orthonormality is with respect to the \Q inner product.}
basis of \emph{graph Fourier modes} $\{\uv_l\}_l$ defined as solution to the following iterative minimizations:
\begin{align*}
  \min_{\uv_L: \forall l<L,\langle\uv_{L},\uv_l\rangle_\Q=0} \Delta(\uv_L) & \quad\text{subj. to}\quad\|\uv_L\|_\Q=1
  \text{.}
\end{align*}
Since $\Delta(\xv)=\xv^*\Mm\xv$, the set of graph Fourier modes that is solution to the minimizations above is exactly the set of eigenvectors of the matrix $\Zm=\Qinv\Mm$ \cite[Theorem 1]{Girault.TSP.2018}.
The analysis and synthesis formula for the $(\Mm,\Q)$-GFT are then given by:
\begin{align*}
  \widetilde{\xv}_l = [\Fm\xv]_l &= \langle\xv,\uv_l\rangle_\Q = [\Um^*\Q\xv]_l\\
  \xv = \Fm^{-1}\widetilde{\xv} &= \Um\widetilde{\xv} =  \sum_l\widetilde{\xv}_l\uv_l=\sum_l\langle\xv,\uv_l\rangle_\Q\uv_l
  \text{,}
\end{align*}
with $\Um=[\uv_0 \dots \uv_{N-1}]$ the matrix whose columns are the graph Fourier modes.
Moreover, the graph frequencies are naturally defined as the graph variations of the graph Fourier modes:
\[
  \lambda_l=\Delta(\uv_l)=\uv_l^*\Mm\uv_l
  \text{,}
\]
and are equal to the eigenvalues of \Zm.

In this setting, a graph signal is $\omega$-bandlimited if:
\[
  \forall \lambda_l>\omega, \widetilde{\xv}_l = 0
  \text{.}
\]

\subsection{Consistent Reconstruction in Hilbert Spaces}

Similar to \cite{Anis.TSP.2016}, our goal is to sample vertices of a graph such that the error made during reconstruction of an $\omega$-bandlimited graph signal is minimal.
In other words, given a signal $\xv$ of which we only know the noisy samples $\yv_\mathcal{S}=\xv_\mathcal{S}+\nv$ on vertices in $\mathcal{S}\subseteq\mathcal{V}$, we would like to recover $\xv$ from $\yv_\mathcal{S}$.
We denote $\widehat{\xv}$ the estimated signal.

Before looking at the question of sampling set selection, we study the problem of signal reconstruction given a sampling set $\mathcal{S}$.
One approach called \emph{consistent reconstruction} is to perform a least squares estimate of the spectral coefficients of the signal and apply the inverse Fourier transform, thus leveraging the bandlimited hypothesis of the signal \xv \cite{Eldar.JFAA.2003}:
\[
  \widehat{\xv}
    = \Um_{\mathcal{V}\mathcal{R}}\left(\argmin_{\widetilde{\xv}_\mathcal{R}} \left\|\yv_\mathcal{S}-\Um_{\mathcal{S}\mathcal{R}}\widetilde{\xv}_\mathcal{R}\right\|_{\QS}^2\right) 
  \text{,}
\]
where $\mathcal{R}=\{1,\dots,r\}$ and $\lambda_r$ is the largest frequency smaller than $\omega$.
Under mild conditions, this estimate verifies $\widehat{\xv}_\mathcal{S}=\yv_\mathcal{S}$.
Notice above the use of the inner product $\|.\|_{\QS}$ of the Hilbert subspace of graph signals supported by the vertex sampling subspace $\mathcal{S}$: This is necessary since $\yv_s$ lives in the Hilbert space of graph signals whose inner product matrix is \QS, not the usual dot product.
We obtain the following close-form solution:%
\footnote{Note that \QS is itself a proper inner product matrix on the graph signals on $\mathcal{S}$ \cite[Observation 7.1.2]{Horn.BOOK.2012}}
\begin{equation}
\label{eq:consistent-reconstruction}
  \widehat{\xv} = \Um_{\mathcal{V}\mathcal{R}}\Bigl(\Um_{\mathcal{S}\mathcal{R}}^*\QS\Um_{\mathcal{S}\mathcal{R}}\Bigr)^{-1}\Um_{\mathcal{S}\mathcal{R}}^*\QS\yv_\mathcal{S}
  \text{.}
\end{equation}

\subsubsection{Effect of Noise}

Here we study the impact of noisy measurements on reconstruction.
More precisely, we look at the error covariance matrix when the noise distribution is a \QS-white noise, \ie, $\nv\sim\mathcal{N}(\zerov,\QSinv)$ \cite{Girault.TSP.2018}.
Considering this type of noise is necessary to have a noise with flat power spectrum, \ie, with $\mathbb{E}[\widetilde{\nv}\widetilde{\nv}^*]=\Im$.
Let $\ev=\widehat{\xv}-\xv$ be this error.
In the Hilbert space of graph signal with inner product matrix \Q, the error covariance matrix is then:%
\footnote{We use the generalization of covariance matrix \Km to arbitrary Hilbert spaces verifying $\langle\Km\xv,\yv\rangle=\mathbb{E}_\zv\left[\langle\zv,\yv\rangle\langle\xv,\zv\rangle\right]$.
Notice also how the covariance matrix of \nv is actually the identity matrix in the corresponding Hilbert space: $\mathbb{E}[\nv\nv^*\QS]=\QSinv\QS=\Im$.}
\[
  \Em = \mathbb{E}\left[\ev\ev^*\Q\right]
      = \Um_{\mathcal{V}\mathcal{R}} \left[\Um_{\mathcal{S}\mathcal{R}}^*\QS\Um_{\mathcal{S}\mathcal{R}}\right]^{-1}\Um_{\mathcal{V}\mathcal{R}}^*\Q
  \text{.}
\]
The A-optimal sampling design minimizing the \emph{\Q-Mean Squared Error} (\Q-MSE) \cite{Girault.TSP.2018} is then obtained as:
\[
  \mathcal{S}^{\text{A-opt}} = \argmin_{|\mathcal{S}|=m} \trace(\Em)
    = \argmin_{|\mathcal{S}|=m} \trace\left(\left[\Um_{\mathcal{S}\mathcal{R}}^*\QS\Um_{\mathcal{S}\mathcal{R}}\right]^{-1}\right)
  \text{,}
\]
while the E-optimal sampling design minimizing the maximum eigenvalue of \Em is given by:
\begin{equation}
\label{eq:e-opt}
  \mathcal{S}^{\text{E-opt}} = \argmin_{|\mathcal{S}|=m} \lambda_{\text{max}}(\Em)
    = \argmax_{|\mathcal{S}|=m} \sigma_{\text{min}}\left(\QSsqrt\Um_{\mathcal{S}\mathcal{R}}\right)
  \text{.}
\end{equation}
In particular, a larger value of $\sigma_{\text{min}}\bigl(\QSsqrt\Um_{\mathcal{S}\mathcal{R}}\bigr)$ leads to a smaller reconstruction error under noisy observations.

\subsubsection{Effect of Model Mismatch}

Now, we assume that the signal we wish to reconstruct is not exactly bandlimited and can be written as $\xv=\xv^\parallel+\xv^\bot$, where $\xv^\parallel$ is the orthogonal projection of \xv on the space of $\omega$-bandlimited signals, and the non-zero $\xv^\bot$ correspond to the high-pass components of the signal.
Using the E-optimal sampling design, the reconstruction error can be bounded by \cite{Eldar.JFAA.2003}:
\begin{equation}
\label{eq:error-bound}
  \left\|\xv-\widehat{\xv}\right\|_\Q \leqslant
    \left[\sigma_{\text{min}}\left(\QSsqrt\Um_{\mathcal{S}\mathcal{R}}\right)\right]^{-1}\left\|\xv^\bot\right\|_\Q
  \text{.}
\end{equation}
In other words, a choice of \Mm and \Q that leads to a larger minimum singular value of $\smash{\QSsqrtsm}\Um_{\mathcal{S}\mathcal{R}}$ leads to a smaller influence of a model mismatch on the reconstruction error.

Noticeably, the influence of both the noise and model mismatch are minimized when $\sigma_{\text{min}}\bigl(\smash{\QSsqrtsm}\Um_{\mathcal{S}\mathcal{R}}\bigr)$ is maximized.
We will use this to experimentally study the performance of various choices of the matrix \Q in \autoref{sec:experiments}.

\subsection{Sampling Set Selection using Spectral Proxies}

In \cite{Anis.TSP.2016}, the authors propose to select a sampling set using:
\begin{equation}
\label{eq:cutoff_sampling_opt}
  \mathcal{S}_k^{\text{opt}} = \argmax_{|\mathcal{S}|=m} \Omega_k(\mathcal{S})
  \text{,}
\end{equation}
where $\Omega_k(\mathcal{S})$ is the $k^{\text{th}}$ order approximate cutoff frequency of the sampling set $\mathcal{S}$, \ie, the minimum bandwidth of a graph signal verifying $\xv_\mathcal{S}=0$.
This cutoff frequency is in turn defined using the $k^{\text{th}}$ spectral proxy of the cutoff frequency of a given graph signal:
\[
  \omega_k(\xv) =
    \raisebox{1ex}{$\displaystyle
      \left(
        \raisebox{-1ex}{$\displaystyle
          \frac{\left\|\left(\Qinv\Mm\right)^k\xv\right\|_{\Q}}{\|\xv\|_{\Q}}
        $}
      \right)^{1/k}
    $}
  \text{,}
\]
leading to:
\begin{align}
\label{eq:Omegak}
  \Omega_k\left(\mathcal{S}\right)
    &= \min_{\xv:\xv_{\mathcal{S}}=\zerov} \omega_k(\xv) \\
    &= \left[\min_{\xv_{\mathcal{S}^c}}
    \frac{{\xv_{\mathcal{S}^c}}^*\Hm_k\left(\mathcal{S}^c\right)^*\Hm_k\left(\mathcal{S}^c\right)\xv_{\mathcal{S}^c}}%
         {{\xv_{\mathcal{S}^c}}^*\QSC\xv_{\mathcal{S}^c}}
    \right]^{\frac{1}{2k}} \nonumber\\
    &= \left[\sigma_{\text{min}}\left(\Hm_k\left(\mathcal{S}^c\right)\right)\right]^{\frac{1}{k}} \nonumber
  \text{,}
\end{align}
where $.^c$ is the set complement operation, with associated singular vector $\phiv_k^*$, and with:
\[
  \Hm_k\left(\mathcal{S}^c\right) = \left[\Qsqrt\left(\Qinv\Mm\right)^k\right]_{\mathcal{V}\mathcal{S}^c}\QSCsqrtinv
  \text{.}
\]

The optimization problem of \autoref{eq:cutoff_sampling_opt} being combinatorial, a greedy heuristic is used to select the vertices of the sampling set one after the other.
More precisely, the cost of adding vertex is obtained using:
\begin{align*}
  \omega_k^\alpha(\xv,\tv)
    &= \left(\omega_k(\xv)+\alpha\frac{\xv^*\diag(\tv)\xv}{\xv^*\xv}\right) \\
  \lambda_k^{\alpha}(\tv)
    &= \min_\xv\omega_k^\alpha(\xv,\tv)
\end{align*}
such that adding vertex $i$ to the sampling set $\mathcal{S}$ leads to:
\[
  \frac{\partial\lambda_k^\alpha}{\partial t_i}(\onev_\mathcal{S})
    = \frac{\partial\omega_k^\alpha}{\partial t_i}(\phiv_k^*,\onev_\mathcal{S})
    = \alpha\left(\frac{\left[\phiv_k^*\right]_i}{\|\phiv_k^*\|}\right)^2
  \text{.}
\]
Notice above how $\smash{\Q}$ does not appear in the norm: The relaxation $\smash{\frac{\xv^*\diag(\onev_\mathcal{S})\xv}{\xv^*\xv}}$ (when $\tv=\onev_\mathcal{S}$) of the constraint $\xv_{\mathcal{S}}=\zerov$ in \autoref{eq:Omegak} does not involve $\smash{\Q}$.
Indeed, the constraint is enforcing that $\xv$ is zero on the set $\mathcal{S}$, which we translate into most of the nodes of $\mathcal{S}$ having values close to zero, rather than the $\QS$-norm of $\xv_{\mathcal{S}}$ being close to zero compared to $\|\xv\|_\Q$.
Note that choosing the $\Q$-norm can be shown to be equivalent to using the $(\smash{\Qsqrtinv\Mm\Qsqrtinv},\Im)$-GFT using the change of variable $\yv=\smash{\Qsqrt\xv}$ in $\lambda_k^\alpha$ which would defeat the purpose of an alternative Hilbert space of graph signals.

We then adopt the same greedy approach that selects the next vertex $i$ that maximizes $|\left[\phiv_k^*\right]_i|$, \ie, that increases the most the cutoff frequency thus optimizing \autoref{eq:Omegak}.

\subsection{Additional Background: Reconstruction with POCS}

Although \autoref{eq:consistent-reconstruction} can be used to reconstruct a graph signal, it requires computing the $(\Mm,\Q)$-GFT matrix, \ie, computing an eigendecomposition.
To avoid performing this eigendecomposition, we adopt the \emph{Projection On Convex Sets} (POCS) approach of \cite{Narang.GLOBALSIP.2013,Gadde.KDD.2014}.
To reconstruct the signal, we iteratively project it onto $(\mathcal{P}_\mathcal{S})$ the set of signals verifying $\widehat{\xv}_\mathcal{S}=\yv_\mathcal{S}$ (recovering the observed samples) and then onto $(\mathcal{P}_\omega)$ the set of $\omega$-bandlimited graph signals until the loop converges or exceeds a maximum number of iterations.

Implementing the projection on $(\mathcal{P}_\omega)$ can be done using an ideal low-pass graph filter of cutoff frequency $\omega$ \cite{Gadde.KDD.2014}.
In practice an ideal low-pass graph filter still needs an eigendecomposition of $\Zm$, so instead, we adopt the same strategy of approximating it with a Chebyshev polynomial approximation of the following graph filter frequency response:
\[
  h(\lambda)=\frac{1}{1 + \exp\left(\alpha(\lambda-\omega)\right)}
  \text{,}
\]
with $\alpha$ controlling how sharp the transition is around the cutoff frequency.
We denote $h_{m,\alpha}^\omega(\lambda)$ this polynomial approximation of order $m$, and the projection on $(\mathcal{P}_\omega)$ of the signal $\xv$ is approximated by $h_{m,\alpha}^\omega(\Qinv\Mm)\xv$.

\section{Experiments with Geometric Graphs}
\label{sec:experiments}

One of the challenges of comparing a spectral method efficacy depending on the graph Fourier transform is to actually measure this efficacy and synthesize ground truth data that is independent of the graph Fourier transforms being compared.
To that end, geometric graphs provide an appealing setting where the underlying Euclidean geometry holds the ground truth.
More precisely, let us consider a portion of a two dimensional space, \eg a square, where a spatially continuous phenomena can be measured.
Such a setting can model a variety of real world examples such as weather readings, geologic earth of the ground, or heat propagation in a metal sheet.
In this section, we experiment with this setting of geometric graphs, and study the application of sampling set seletion and signal reconstruction described in \autoref{sec:method}.

There are however a number of parameters that we need to choose before performing vertex sampling:
\begin{inlinelist}
  \item the edge set $\mathcal{E}$
  \item the weights of each edge $w_{ij}$
  \item the graph signal variation operator $\Delta(\xv)$
  \item the graph signal inner product \Q
\end{inlinelist}.
Here we use the classical approach of a complete graph with edge weights given by a Gaussian kernel of the distance between vertices.
Let $\dist(i,j)$ be the Euclidean distance between any two vertices (locations), $w_{ij}=\exp\bigl(-\dist(i,j)^2/(2\sigma^2)\bigr)$ be a Gaussian kernel of the distance, and $\Dm=\diag(\Wm\onev)$ be the degree matrix associated to the weight matrix \Wm.
Then, using the combinatorial Laplacian $\Mm=\Lm=\Dm-\Wm$, we choose the classical graph signal variation $\Delta(\xv)=\xv^*\Lm\xv$.

Using this setting, we compare three $(\Lm,\Q)$-GFTs: $\Q=\Im$, $\Q=\Dm$, and $\Q=\Cm$ where \Cm is the diagonal matrix with Voronoi cell areas of each vertex \cite{Girault.TSP.2018}.
In each experiment, multiple realizations of geometric graphs of 100 vertices are obtained by uniformly drawing vertices in a $10\times 10$ square.

In all of our experiments, we use $k=3$ for the spectral proxies $\omega_k(\xv)$.

\subsection{Error Bound}

\begin{figure}
  \centering
  \includegraphics{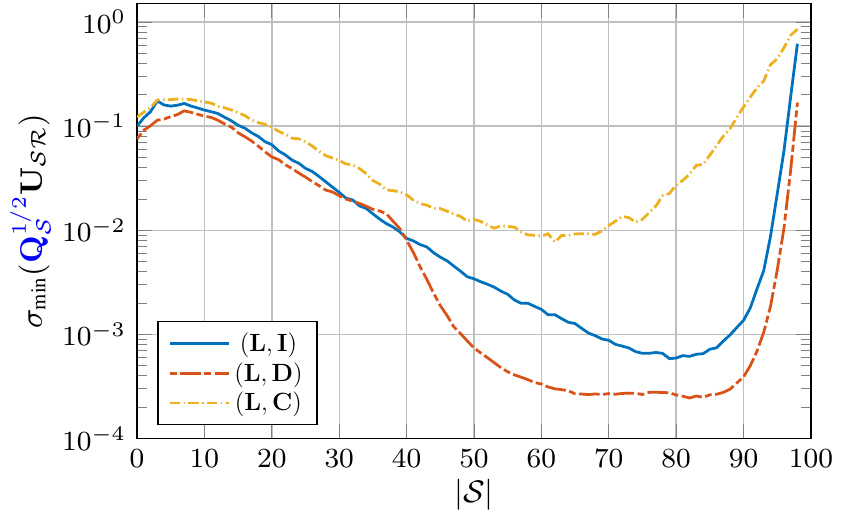}
  
  \caption{Average value of $\sigma_{\text{min}}(\smash{\QSsqrt}\Um_{\mathcal{S}\mathcal{R}})$ over 5,000 realizations of geometric graphs, where $\mathcal{S}$ has been obtained by the method described in \autoref{sec:method}.}
  \label{fig:error-bound}
\end{figure}

In \autoref{eq:error-bound}, we showed that the reconstruction error is upper bounded by a quantity that depends on the minimum singular value of $\smash{\QSsqrtsm}\Um_{\mathcal{S}\mathcal{R}}$, and showed that this minimum singular value should be large to have a smaller worse case reconstruction error.
In \autoref{fig:error-bound}, we show the resulting average minimum singular value for 5,000 realizations of geometric graphs.
$\Q=\Cm$ outperforms the other two choices of inner product matrix \Q.
Interestingly, the random walk Laplacian is the less performing of all three approaches and sees even a large drop of performance for sampling sets larger than 40\% of the vertex set.
This may suggest that this approach is better at capturing global trends on the graph (smooth graph signals easily described by a small subset of vertices) than finer grained details.
However, the reconstruction error also depends on the model mismatch energy (second term in the r.h.s. of \autoref{eq:error-bound}).
To better quantify this reconstruction error, we study it on concrete signals next.

\begin{figure*}[t]
  \centering
  
  \includegraphics{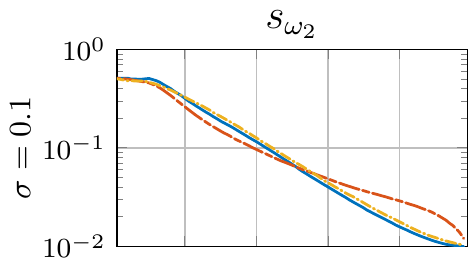}
  \includegraphics{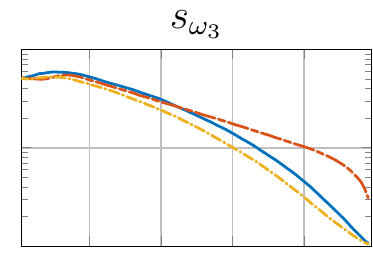}
  \includegraphics{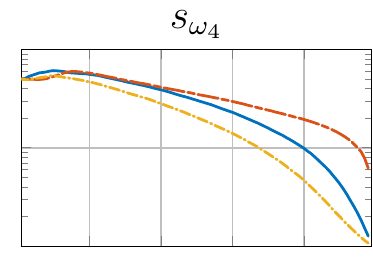}
  \includegraphics{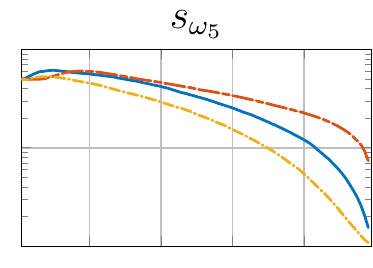}
  
  \includegraphics{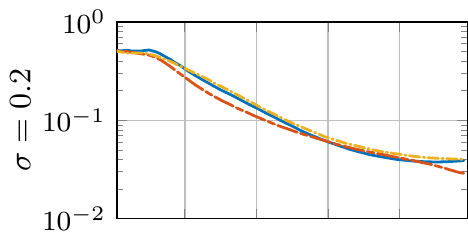}
  \includegraphics{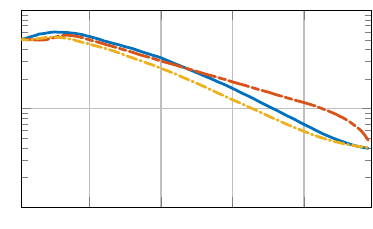}
  \includegraphics{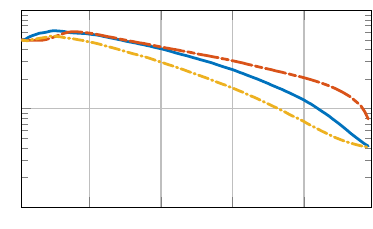}
  \includegraphics{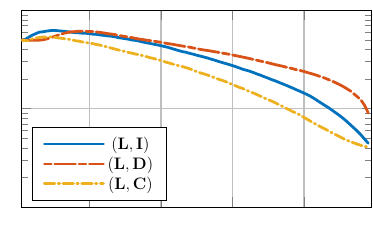}
  
  \includegraphics{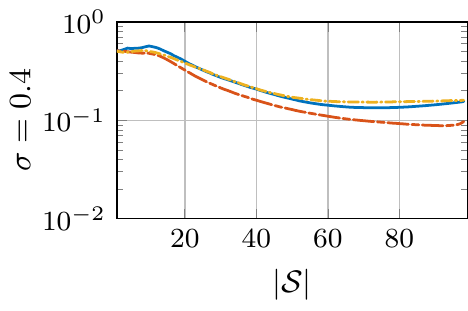}
  \includegraphics{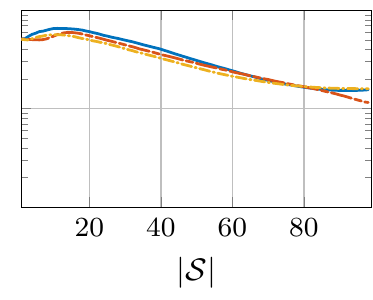}
  \includegraphics{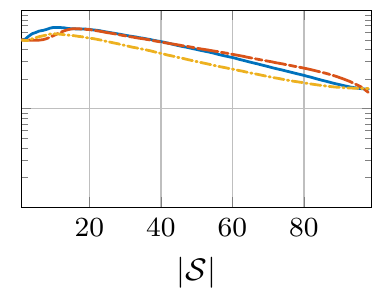}
  \includegraphics{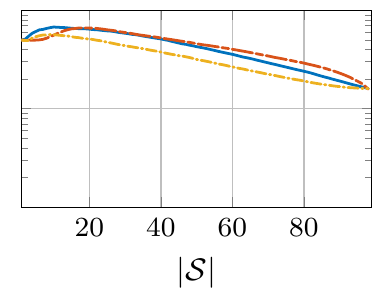}
  
  \caption{Mean $\Cm$-norm of the reconstruction error for various ground truth graph signals $s_{\omega_k}$ (columns), and various amount of additive Gaussian noise with standard deviation $\sigma$ (rows), depending on the size of the sampling set $|\mathcal{S}|$.}
  \label{fig:mse}
\end{figure*}

Although this only shows that the worse case reconstruction scenario is less impacted by noise with $\Q=\Cm$, this highlights the property that choosing a quantity that more finely describes the underlying structure can actually help in making the right sampling set selection.

\subsection{Synthetic Graph Signals}

In this section, we are interested in simulating sampling and reconstruction of graph signals, and studying the average error made by all three choices of matrix \Q.
Noticeably, the usual approach of choosing a bandlimited ground truth graph signal and perturbing it with additive Gaussian noise is flawed in our setting: There are multiple definitions of bandlimitedness that each depend on a different $(\Lm,\Q)$-GFT.
To lift this difficulty, and to be true to our motivations, we consider a smooth ground truth continuous signal.
More precisely, we consider four pure horizontal sinewaves $s_\omega(x,y)=\sin(2\pi\omega x)$ and choose $\omega=\omega_k$ such that there are exactly $k\in\{2,3,4,5\}$ horizontal oscillations in the underlying continuous space.
To study the mean squared error, we also perturb the measurements of $s_\omega$ at each vertex with a Gaussian noise $n_i\sim\mathcal{N}(0,\sigma^2)$, independently for each vertex.

Note that this additive noise corresponds to measurement errors.
Even though it is modeled differently depending on \Q, choosing one \Q over another is not about filtering out uniformly the noise, but filtering it out where it is important to do so.
In other words, using an approach that removes uniformly the noise ($\Q=\Im$) may be detrimental if the noised removed is not important, while selectively removing the noise can help attain our goal (here, having an accurate representation of the underlying continuous signal).

Additionally, we do not plot the $\ell_2$-norm of the error, but the $\Cm$-norm, based on the Voronoi cell area since we showed previously that it corresponds to the error made in the underlying continuous space if we were interpolating the measurement in the continuous domain \cite{Girault.TSP.2018}.
It is therefore a more realistic measure of how accurate our reconstruction is with respect to the underlying continuous space.

\autoref{fig:mse} reports these mean squared errors for each of the 12 cases (four signals, and three noise levels with standard deviations $\sigma\in\{0.1,0.2,0.4\}$).
Noticeably, the $(\Lm,\Dm)$-GFT based approach slightly outperforms the other two for the smoothest signal, but quickly becomes less effective as signal smoothness decreases.
For $s_{\omega_3}$ and $s_{\omega_4}$ we remark that the MSE of the $(\Lm,\Dm)$-GFT approach actually decreases for a larger sampling set size than the other two methods.
As stated in \cite{Anis.TSP.2016}, this suggests that the cutoff frequency of the signal being considered is actually higher with respect to the $(\Lm,\Dm)$-GFT than the other two methods.

On the other hand, $(\Lm,\Cm)$ performed the best for $s_{\omega_3}$, $s_{\omega_4}$ and $s_{\omega_5}$ with barely larger error for smaller sampling sets, and then outperforming the other two methods, thus showing the importance of correctly choosing \Q.

\section{Conclusions and Perspectives}

In this paper, we showed how to extend the sampling set selection through spectral proxies method of \cite{Anis.TSP.2016} to any Hilbert space of graph signals.
Doing so, we highlighted how the inner product matrix \Q impacts the method, and the assumptions made to derive close-form solutions.
We also showed on synthetic geometric graphs experiments how the choice of \Q alters reconstruction error bounds and reconstruction errors for some smooth graph signals.

This work opens up several interesting perspectives.
In particular, the Voronoi cell area inner product is shown to be helpful, but can we do better?
Can we find a better inner product for geometric graphs, or an alternative choice of edge weights to better match the inner product?
This also opens up many possibilities for other types of graphs, such as social network graphs for which we know already that the heavy-tail degree distribution is an obstacle to graph signal processing.

\bibliographystyle{IEEEbib}
\bibliography{bibliography}

\begin{thebibliography}{10}

\bibitem{Shuman.SPMAG.2013}
David~I. {Shuman}, Sunil~K. {Narang}, Pascal {Frossard}, Antonio {Ortega}, and
  Pierre {Vandergheynst},
\newblock ``{The emerging field of signal processing on graphs: Extending
  high-dimensional data analysis to networks and other irregular domains},''
\newblock {\em IEEE Signal Processing Magazine}, vol. 30, no. 3, pp. 83--98,
  May 2013.

\bibitem{Ortega.PROCIEEE}
Antonio {Ortega}, Pascal {Frossard}, Jelena {Kova{\v c}evi{\'c}}, Jos{\'e}
  M.~F. {Moura}, and Pierre {Vandergheynst},
\newblock ``{Graph Signal Processing: Overview, Challenges, and
  Applications},''
\newblock {\em Proceedings of the IEEE}, vol. 106, no. 5, pp. 808--828, May
  2018.

\bibitem{Shomorony.GLOBALSIP.2014}
Ilan Shomorony and A.~Salman Avestimehr,
\newblock ``{Sampling large data on graphs},''
\newblock in {\em {2014 IEEE Global Conference on Signal and Information
  Processing (GlobalSIP)}}. Dec. 2014, pp. 933--936, IEEE.

\bibitem{Chen.TSP.2015}
Siheng Chen, Rohan Varma, Aliaksei Sandryhaila, and Jelena Kova{\v c}evi{\'c},
\newblock ``{Discrete Signal Processing on Graphs: Sampling Theory},''
\newblock {\em {IEEE Transactions on Signal Processing}}, vol. 63, no. 24, pp.
  6510--6523, Dec. 2015.

\bibitem{Anis.TSP.2016}
Aamir Anis, Akshay Gadde, and Antonio Ortega,
\newblock ``{Efficient Sampling Set Selection for Bandlimited Graph Signals
  Using Graph Spectral Proxies},''
\newblock {\em {IEEE Transactions on Signal Processing}}, vol. 64, no. 14, pp.
  3775--3789, July 2016.

\bibitem{Girault.TSP.2018}
Benjamin Girault, Antonio Ortega, and Shrikanth~S. Narayanan,
\newblock ``{Irregularity-Aware Graph Fourier Transforms},''
\newblock {\em {IEEE Transactions on Signal Processing}}, vol. 66, no. 21, pp.
  5746--5761, Nov. 2018.

\bibitem{vonLuxburg.SC.2007}
Ulrike von Luxburg,
\newblock ``{A tutorial on spectral clustering},''
\newblock {\em Statistics and Computing}, vol. 17, no. 4, pp. 395--416, Dec
  2007.

\bibitem{Milanfar.SPMAG.2013}
Peyman Milanfar,
\newblock ``{A Tour of Modern Image Filtering: New Insights and Methods, Both
  Practical and Theoretical.},''
\newblock {\em IEEE Signal Process. Mag.}, vol. 30, no. 1, pp. 106--128, 2013.

\bibitem{Lu.ICIPPRE.2020}
Keng-Shih Lu, Antonio Ortega, Debargha Mukherjee, and Yue Chen,
\newblock ``{Perceptually Inspired Weighted MSE Optimization using
  Irregularity-Aware Graph Fourier Transform},''
\newblock submitted to ICIP 2020.

\bibitem{deQueiroz.TIP.2016}
Ricardo~L. de~Queiroz and Philip~A. Chou,
\newblock ``{Compression of 3D Point Clouds Using a Region-Adaptive
  Hierarchical Transform},''
\newblock {\em {IEEE Transactions on Image Processing}}, vol. 25, no. 8, pp.
  3947–3956, Aug. 2016.

\bibitem{Chou.TIP.2020}
Philip~A. Chou, Maxim Koroteev, and Maja Krivokuća,
\newblock ``{A Volumetric Approach to Point Cloud Compression—Part I:
  Attribute Compression},''
\newblock {\em {IEEE Transactions on Image Processing}}, vol. 29, pp.
  2203–2216, 2020.

\bibitem{Pavez.ICIPPRE.2020}
Eduardo Pavez, Benjamin Girault, Antonio Ortega, and Philip~A. Chou,
\newblock ``{Region Adaptive Graph Fourier Transform for 3d Point Clouds},''
\newblock submitted to ICIP 2020.

\bibitem{Eldar.JFAA.2003}
Yonina~C. Eldar,
\newblock ``{Sampling with Arbitrary Sampling and Reconstruction Spaces and
  Oblique Dual Frame Vectors},''
\newblock {\em {Journal of Fourier Analysis and Applications}}, vol. 9, no. 1,
  pp. 77--96, 2003.

\bibitem{Horn.BOOK.2012}
Roger~A. Horn and Charles~R. Johnson,
\newblock {\em {Matrix Analysis}},
\newblock Number 2nd Edition. Cambridge University Press, oct 2012.

\bibitem{Narang.GLOBALSIP.2013}
Sunil~K Narang, Akshay Gadde, Eduard Sanou, and Antonio Ortega,
\newblock ``{Localized iterative methods for interpolation in graph structured
  data},''
\newblock in {\em {2013 IEEE Global Conference on Signal and Information
  Processing}}. Dec. 2013, pp. 491--494, IEEE.

\bibitem{Gadde.KDD.2014}
Akshay Gadde, Aamir Anis, and Antonio Ortega,
\newblock ``{Active Semi-Supervised Learning using Sampling Theory for Graph
  Signals},''
\newblock in {\em {20th ACM SIGKDD International Conference on Knowledge
  Discovery and Data Mining}}. ACM, Aug. 2014, pp. 492--501, ACM.

\end{thebibliography}

\end{document}